\documentclass[aps,floatfix, twocolumn]{revtex4} 

\usepackage{graphicx}
\usepackage{dcolumn}
\usepackage{bm}
\usepackage{longtable}
 \usepackage{amsmath}
 \usepackage{amssymb}
 \usepackage{eurosym}
\usepackage{float}
\begin{document}

\title{Systemic trade-risk of critical resources}

\author{Peter Klimek$^1$, Michael Obersteiner$^2$, Stefan Thurner$^{1,2,3,*}$}
\email{stefan.thurner@meduniwien.ac.at}

\affiliation{$^1$Section for Science of Complex Systems; Medical University of Vienna; 
Spitalgasse 23; A-1090; Austria\\ 
$^2$IIASA, Schlossplatz 1, A 2361 Laxenburg; Austria\\
$^3$Santa Fe Institute; 1399 Hyde Park Road; Santa Fe; NM 87501; United States
}

\begin{abstract} 
In the wake of the 2008 financial crisis the role of strongly interconnected markets in fostering systemic instability has been increasingly acknowledged.
Trade networks of commodities are susceptible to deleterious cascades of supply shocks that increase systemic trade-risks and pose a threat to geopolitical stability.
On a global and a regional level we show that supply risk, scarcity, and price volatility of non-fuel mineral resources are intricately connected with the structure of the world-trade network of or spanned by these resources.
On the global level we demonstrate that the scarcity of a resource, as measured by its trade volume compared to extractable reserves, is closely related to the susceptibility of the trade network with respect to cascading shocks.
On the regional level we find that to some extent the region-specific price volatility and supply risk can be understood by centrality measures that capture systemic trade-risk.
The resources associated with the highest systemic trade-risk indicators are often those that are produced as byproducts of major metals.
We identify significant shortcomings in the management of systemic trade-risk, in particular in the EU.
 \end{abstract}
 
\maketitle
\paragraph*{}
Commodity price volatility has long been identified by political economists as a hindrance for sustainable economic development (e.g., the Dutch Disease) as well as a catalyst and product of geopolitical crises.
Although traditionally associated with fossil fuel resources, the criticality of non-fuel mineral resources is increasingly relevant due to their use in cutting edge technological and medical applications \cite{GraedelPNAS}. 
With the explosive growth of financial derivatives on commodities and a subsequent investment boom and bust in the mid 2000s, there is growing evidence that resource criticality, loosely defined as the importance of a resource to production processes, is increasingly susceptible to financial perturbations both, from within and outside the commodities sector \cite{tang2010index, baffes2010placing}.
A better understanding of the interconnected nature of commodity markets would allow policymakers to hedge against threats to industrial sectors and to reduce risk of geopolitical instabilities that are induced by resource price volatility.

Systemic risk is often defined as the risk that a large fraction of a system collapses as a consequence of seemingly minor and local shocks that initially only affect a small part of the system.
Due to the interconnectedness of the system these shocks may cause secondary effects that eventually propagate through the entire network.
Awareness of systemic risk has greatly increased in the finance literature in the wake of the 2008 financial crisis \cite{billio2012econometric, huang2009framework}.
For financial systems it has been shown that systemic risk is to a large extent a network effect \cite{haldane2011systemic}.
There external shocks to a single financial institution may result in a sudden reduction of financial flows to other institutions which causes distress for them as well.
This chain of reduced financial flows can spread through the system, can lead to positive feedback dynamics, and may end in a strong reduction of the total net worth of financial institutions \cite{Calvo2008}.
It has been shown that the vulnerability of a system to such cascading shocks can be assessed by network centrality measures and related concepts \cite{grubesic2008comparative, simonsen2008transient, Battiston12, ThurnerSRT}.

It is becoming increasingly clear that security of supply can only be understood in a framework that acknowledges the global interconnections among resource systems \cite{Chatham, erdmann2011criticality}.
In this work we show that the likelihood of price disruptions in mineral prices is strongly related to the structure of the trade network of the particular resource.
We introduce a novel method to assess the systemic risk of trade networks and demonstrate its validity on the actual trade networks of 71 resources.
On a global level we show that the scarcity of a resource is strongly related to structural properties of the trade networks.
The scarcer the resource, the more susceptible to cascading shocks is the trade network.
On a regional level we show that the volatility of mineral prices within several world regions, in particular the US and the EU, is closely related to specific network centrality measures that we design to capture systemic trade-risk.
We find that price disruptions of mineral resources also reflect cascades of supply shocks in the underlying trade network.
The impact of these cascades can to some extent be mitigated by lowering trade barriers.
We find the highest systemic trade-risk for resources that are primarily produced as byproducts of other resources.
It has been argued that these resources are especially prone to price disruptions since it is hard to predict whether their global supply will react to changes in global demand \cite{Graedel14}.

\begin{figure*}[tbp]
\begin{center}
 \includegraphics[width=130mm]{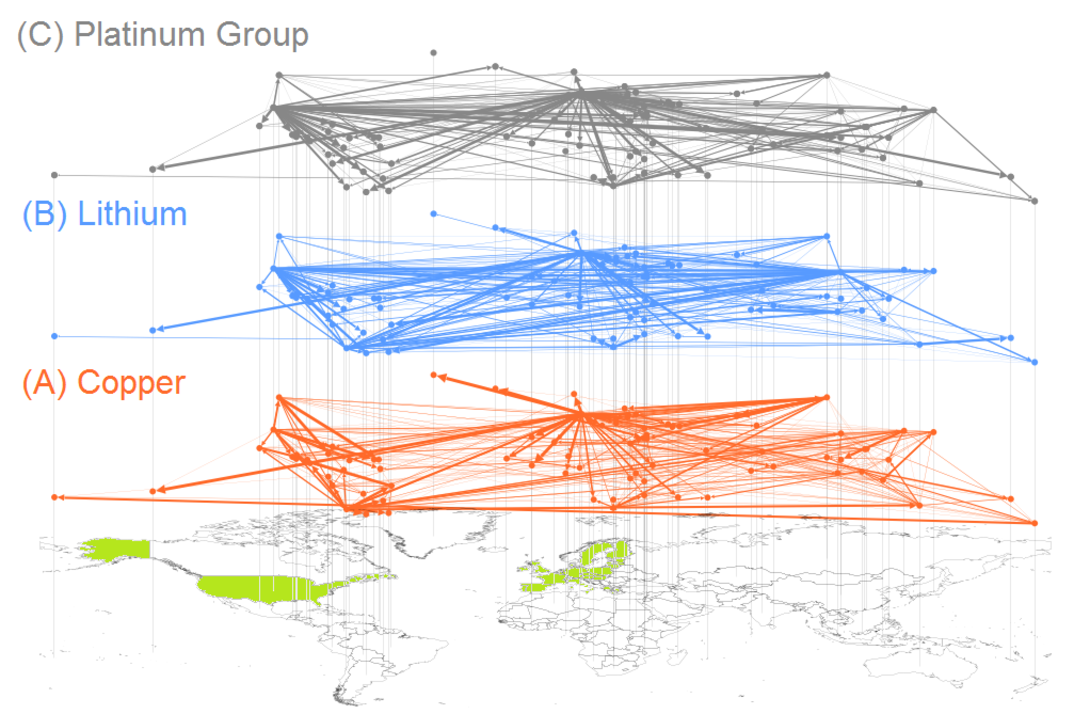}
\end{center}
 \caption{The world-wide trade-risk network for non-fuel minerals, represented as a multiplex trade network $V_{ij}^{r}(t)$, where each layer corresponds to one mineral resource, (A) copper, (B) lithium, and (C) platinum group metals. We study the network topology of each of these layers and compute both regional (node-based, i.e. country-specific) and global (network-based) measures. We study the relationships between supply risk, price volatility, network centrality, and trade barriers for the US and EU (world regions highlighted in green in the world map).}
 \label{networks}
\end{figure*} 

\section{`TradeRisk': an indicator for systemic trade-risk}

We work with 71 non-fuel mineral resources as provided by the US Geological Survey (USGS) in the Mineral commodity summaries \cite{USGS}.
For each of these resources ($r$) we construct the network of international cross-border trade flows $M_{ij}^{r}(t)$.
The result is a so-called multiplex network where nodes ($i$, $j$) represent countries that are connected by different types of links ($r$) that represent trade in different commodities.
The entries in $M_{ij}^{r}(t)$ represent the amount of resource $r$ in USD that flows from country $i$ to country $j$ within year $t$.
Details on how $M_{ij}^{r}(t)$ is extracted from the data are discussed in section \ref{sec: mtd}.

The vulnerability to supply shocks in mineral imports of countries has a strong geopolitical component.
Imports from countries that are politically unstable are more prone to supply restrictions than imports from countries that are politically stable \cite{Graedel}.
The indicator `Political Stability and Absence of Violence', $PS_i(t)$, measures the likelihood of political, social or economic distress in country $i$ in year $t$ \cite{WGI}.
$PS_i(t)$ ranges from zero to hundred.
High values indicate high political stability. 
The network-based vulnerability of country $j$ to import shortages of mineral $r$, due to supply restrictions from another country $i$, is given by the trade-risk multiplex network $V_{ij}^{r}(t)$, defined as
\begin{equation}
V_{ij}^{r}(t) = \left( 1-\tfrac{PS_i(t)}{100} \right) \frac{M_{ij}^{r}(t)}{\sum_i M_{ij}^{r}(t)} \quad.
\label{vuln}
\end{equation}
$V_{ij}^{r}(t)$ is the fraction of country $j$'s imports of commodity $r$ from $i$, in year $t$, weighted by how likely country $i$ experiences political or social disturbances.
The trade-risk vulnerability network $V_{ij}^{r}(t)$ is shown for copper, lithium, and platinum group metals in figure \ref{networks}.
As an alternative to the Political Stability indicator we also use the Resource Governance Index $RGI_i$ instead of $PS_i(t)$.
$RGI_i$ measures the quality of governance in the oil, gas and mining sectors on a scale from zero to hundred \cite{RGI}.

Imagine a country that receives its imports from a large number of politically stable countries, but those countries all rely on imports from a single, politically unstable country.
Clearly, any measure for supply risk that is only based on trade flows with direct neighbors in the trade network does not take such situations into account.
These influences can be quantified by recursive centrality measures, for example the PageRank \cite{Newman}.
The PageRank $PR_i^{r}(t)$ of country $i$ in the trade-risk network for resource $r$ at time $t$ is given by solutions to the recursive equation
\begin{equation}
PR_i^{r}(t) = \alpha \sum_j \frac{1}{k_{out,j}^{r}(t)} V_{ij}^{r}(t) PR_j^{r}(t) + (1-\alpha) \quad,
\label{pagerank}
\end{equation}
of the PageRank, where $k_{out,j}^{r}(t)$ is the out-degree (number of countries into which $r$ is exported) of $j$.
Those countries pass the shock on to the countries that receive exports from them, and so on.
The parameter $(1-\alpha)$ can be understood as the contribution to supply shocks due to effects that are not related to the trade-risk network.
Note that equation \ref{pagerank} only converges for $\alpha<\tfrac{1}{\lambda^{r}(t)}$, where $\lambda^{r}(t)$ is the largest eigenvalue of $V^{r}$.
We adopt the standard convention by setting $\alpha \equiv \alpha(r,t) = \tfrac{0.85}{\lambda^{r}(t)}$ \cite{Newman}.
Here we then take a time average $PR_i^{r} =\langle PR_i^{r}(t) \rangle_t$, where $\langle \cdot \rangle_t$ denotes the average over the years 2000-2012.
$PR_i^{r}$ is a measure for how likely country $i$ is affected by supply shocks in any other country, even when there is no direct trade relation between those countries.
A potential shock in country $j$ will be distributed in units of $\tfrac{1}{k_{out,j}^{r}(t)}$ to all countries that receive exports from $j$, where 

Countries will only be vulnerable to changes in their imports of mineral $r$ if they have a non-zero import reliance $IR_i^{r}$.
The $IR_i^{r}$ quantifies how strongly the economy in country $i$ depends on imports of resource $r$, see section \ref{sec: sr}.
Finally, we arrive at the network-based impact of supply shocks in resource $r$ for country $i$, the TradeRisk $TR_i^{r}$, which is given by
\begin{equation}
TR_i^{r} = PR_i^{r} \ IR_i^{r} \quad.
\label{traderisk}
\end{equation}

For each network layer in $V_{ij}^{r}(t)$ we study the following indicators, averaged over the years 2000-2012.
The average degree $\bar k^{r}$ is the average number of non-zero links per node for a given resource $r$.
$SCC^{r}$ is the number of nodes that are part of the largest strongly connected component, divided by the number of nodes in the network.
The SCC is the largest subset of nodes where each node can be reached on the network from every other node.
The largest eigenvalue, $\lambda^{r}$, of $V^{r}$ is a measure for how susceptible the trade-risk network $V^{r}$ is to epidemic spreading processes.
The smaller $\lambda^{r}$, the easier it is for a small shock to propagate through the entire network \cite{Wang03}.
In this sense $1/\lambda^{r}$ can be seen as a measure for resilience of the network.
The scarcity $S^{r}$ of a commodity $r$ is defined as the logarithmic quotient of the total trade volume and the estimated reserves, $S^{r} = \log (TTV^{r}/R^{r})$.
Results for several indicator values for individual resources are given in the supplement in table S.\ref{details}.

We define the adjacency multiplex $B_{ij}^{r}(t) = 1$, if $V_{ij}^{r}(t) > 0$ from equation \ref{vuln}, and $B_{ij}^{r}(t) = 0$ otherwise.
The in-degree of country $j$, $k_{in, j}^{r}$, is given by $k_{in, j}^{r} = \langle \sum_i B_{ij}^{r}(t) \rangle_t$.
$k_{in, j}^{r}$ is the number of countries that contribute to at least one percent of $i$'s total imports of mineral $r$, averaged over all available years.
The in-strength $w_{in, j}^{r}$ for country $j$ is given by $w_{in, j}^{r} = \langle \sum_i V_{ij}^{r}(t) \rangle_t$.
Note that $w_{in, j}^{r}$ can be seen as a weighted average of the political stability of the countries that export $r$ to $i$.
The weights are the fractions of $i$'s total imports in $r$ that the particular countries $j$ provide.
We consider an alternative formulation of the TradeRisk indicator by replacing the PageRank $PR_{i}^{r}$ with the in-strength $w_{in,i}^{r}$ in equation \ref{vuln}, the In-Strength TradeRisk $TR_{i}^{str,(r)} = w_{in,i}^{r} IR_{i}^{r}$.

For testing our results for significance of network effects, we will employ several randomized versions of the data, see also section \ref{sec: rds}.
In the first randomization, $M^{r}_{\textrm{fix degree}}(t)$, we keep the average degree $\bar k^{r}$ fixed and each trade flow gets a randomly selected importing and exporting country. 
The second randomization, $M^{r}_{\textrm{fix in-deg}}(t)$, randomizes the exporting country for each trade flow, but keeps the importing country fixed.
In the third randomization, $M^{r}_{\textrm{fix in-/out-deg}}(t)$, the importing and exporting countries are fixed but randomly permutes the values of the non-zero trade flows.

\section{Results}

\subsection{Global results: resilience and trade networks}

We find that the composite supply risk $CSR^{r}$ has a weak negative correlation with the largest eigenvalue $\lambda^{r}$ of $V^{r}$ (Pearson correlation coefficient $\rho = -0.32$, $p$-value of $p=0.026$), see table \ref{xcorr_glob}.
$CSR^{r}$ is also negatively correlated with the size of the strongly connected component, $SCC^{r}$ ($\rho=-0.41$, $p=0.0039$).
A high production concentration may indicate a small strongly connected component and consequently an increased supply risk.
Both, the largest eigenvalue $\lambda^{r}$ and the $SCC^{r}$ show a significant correlation with the scarcity $S^{r}$.
The scarcer a resource, the less resilient the trade-risk network to supply shocks, and the higher the largest eigenvalue $\lambda^{r}$ ($\rho = 0.47$, $p=0.0011$).
These correlations are not confounded by the influence of the trading volume, $TTV^{r}$, itself, as seen by the non-significant correlations of $\lambda^{r}$ and $SCC^{r}$ with $TTV^{r}$.
The logarithmic average degree $\log \bar k^{r}$ has only a negative correlation with resource scarcity ($\rho=0.31$, $p=0.041$).

Results for the supply risk $CSR^{r}$, scarcity $S^{r}$, and trade volume $TTV^{r}$ for the randomized trade networks $M^{r}_{\textrm{fix degree}}(t)$ are shown in the supporting information in table S.\ref{si_glob}.
By construction the correlations of the average degree with both, the supply risk and scarcity of a resource are preserved under this randomization, see table S.\ref{si_glob}.
However, there is {\it no} significant correlation of the largest eigenvalue $\lambda^{r}$ with the supply risk $CSR^{r}$ or the scarcity $S^{r}$, respectively, in the randomized data.
This shows that resilience to cascading shocks as observed in the real data is indeed a genuine network effect that can not be explained by the size or number of trade flows alone.

\begin{table}[bp]
\caption{Global properties of the trade networks for each resource $r$. The elements in the table the Pearson correlation coefficients. The composite supply risk $CSR^{r}$ is negatively correlated with the largest eigenvalue $\lambda^{r}$ and the size of the SCC. $SCC^{r}$ is positively correlated with both the total trading volume and the scarcity of the resource. The higher the scarcity of the mineral, the lower the resilience to shocks of the trade-risk network. These correlations can not be explained by a potentially confounding influence of the trade volume itself, as seen by the non-significant correlations of $\lambda^{r}$ and $SCC^{r}$ with $TTV^{r}$. We indicate significance, $p<0.05$, by $(^*)$, $p<0.01$ by $(^{**})$, and $p<0.001$ by $(^{***})$.}
\label{xcorr_glob}
\begin{tabular}{l l l l}
\hline
Correlation coefficient &    $CSR^{r}$ &  $S^{r}$  & $TTV^{r}$ \\
\hline
Largest Eigenvalue $\lambda^{r}$ &$-0.32^*$ & $0.47^{**} $& 0.21\\
Size SCC $SCC^{r}$ &$-0.41^{**}$ &$0.45^{***}$ &0.05 \\
\hline
\end{tabular}
\end{table}

\begin{table*}[tbp]
\caption{Regional results for the correlations of the TradeRisk indicators, price volatilities, and trade barriers. Price volatility of mineral resources is best explained using the TradeRisk indicator for both the EU and US. There are also significant correlations between price volatility and import reliance, PageRank, and In-Strength TradeRisk, respectively. The level of applied protection (trade barriers) $TB_i^{r}$ is negatively correlated with TradeRisk in the US, but not in the EU.}
\label{xcorr_loc}
\begin{tabular}{l | l l | l l}
\hline
Correlation with &    $\sigma_{EU}^{r}$  &      $\sigma_{US}^{r}$   & $TB_{EU}^{r}$ & $TB_{US}^{r}$ \\
\hline
TradeRisk $TR_i^{r}$ & $0.71^{***}$ & $0.58^{***}$ & $-0.11$ & $-0.39^{**}$\\
Import Reliance $IR_i^{r}$ & $0.48^{**}$ & $0.51^{***}$ & $-0.15$ & $-0.10$ \\
PageRank $PR_i^{r}$ &$0.56^{***}$ &$0.45^{***}$ & $-0.23$  & $-0.43^{***}$  \\
In-Strength TradeRisk $TR_i^{str,r}$ & $0.39^{*}$ &$0.50^{***}$  & $-0.12$ & $-0.11$ 
\end{tabular}
\end{table*}
\subsection{Region-specific results: price volatility and network effects}

\begin{figure*}[tbp]
\begin{center}
 \includegraphics[width=130mm]{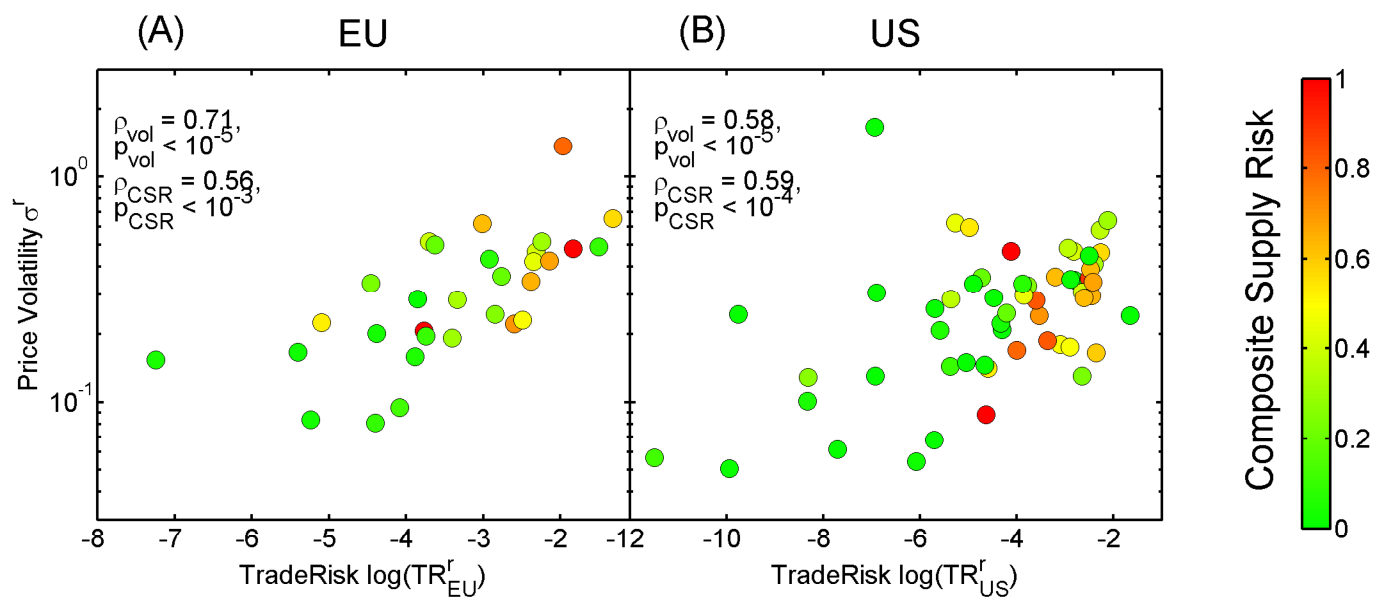}
\end{center}
 \caption{TradeRisk versus price volatility for the EU and the US. Each point represents a mineral resource. The country-specific TradeRisk indicator for (A) EU and (B) US  is significantly correlated with both the average yearly price volatility of the specific mineral, and with the composite supply risk, indicated by color. Resources with high $CSR^{r}$ tend to be on the right hand side. We also show the correlation coefficients $\rho_{\sigma}$ and $\rho_{CSR}$ of the price volatility with TradeRisk and composite supply risk, respectively, together with the $p$-values to reject the null hypothesis that the true correlation coefficient is zero.}
 \label{results2}
\end{figure*}

Region-specific results are computed for the EU and US.
Results for the EU are obtained by condensing the 25 EU members of 2012 into a single node and by computing the TradeRisk, $TR_{EU}^{r}$, in the so-obtained network.
There is a highly significant correlation between price volatility of the resource $r$ in the EU, $\sigma_{EU}^{r}$, and TradeRisk ($\rho=0.71$, $p<10^{-4}$), see figure \ref{results2}(A).
This correlation is a genuine network effect.
To show this we consider an alternative formulation of the TradeRisk indicator by replacing $TR_{EU}^{r}$ with the In-Strength TradeRisk $TR_{EU}^{str,r}$
Table \ref{xcorr_loc} shows that the TradeRisk $TR_{EU}^{r}$ has a higher correlation with price volatility than it has with any of the other indicators, namely the import reliance $IR_{EU}^{r}$, the PageRank $PR_{EU}^{r}$, and the In-Strength TradeRisk $TR_{EU}^{str,r}$.
To understand the impact of higher-order network effects on volatility of resource prices, we study the linear partial correlation, $\rho_{partial}$, between $TR_{EU}^{r}$ and $\sigma_{EU}^{r}$, controlling for the influence of  $TR_{EU}^{str,r}$.
The partial correlation $\rho_{partial}$ can be interpreted as the amount of variance in $\sigma_{EU}^{r}$ that can only be explained by knowledge of the entire trade-risk network, after the influence of direct neighbors in the network has been removed.
We find that $\rho_{partial} = 0.68$ ($p<10^{-4}$), which means that about $96\%$ of the original correlation between price volatility and TradeRisk (which was $\rho = 0.71$) can be attributed to genuine network effects.

Basically the same observations also hold for the US, see figure \ref{results2}(B),
where again the TradeRisk indicator explains price fluctuations, $\sigma_{US}^{r}$ ($\rho=0.58$, $p<10^{-5}$), better than the In-Strength TradeRisk, the import reliance, and the PageRank alone.
After controlling for the influence of the In-Strength TradeRisk $TR_{US}^{str,r}$, we find a partial correlation of $\rho_{partial} = 0.38$ ($p=0.0032$) between TradeRisk and price volatility.
In both regions, EU and US, there is a significant correlation between TradeRisk and supply risk $CSR^{r}$, see figure \ref{results2}.
This result is not surprising, since both indicators explicitly depend on the import reliance and the political stability of the top-producing countries.

Table \ref{xcorr_loc} shows that there are significant differences between the US and the EU in the correlations of TradeRisk with the level of protection, the trade barriers $TB_i^{r}$.
The US tends to employ lower trade barriers for the imports of resources with high systemic trade-risk, whereas there is no significant relation between TradeRisk and $TB_{EU}^{r}$ in the EU.
Table \ref{xcorr_loc} also shows that the high correlation between TradeRisk and $TB_{US}^{r}$ is driven by the PageRank contributions to systemic trade-risk, which shows that the US have lower barriers for resources where they have a high network-based vulnerability (and not necessarily a high import reliance).

Replacing the political stability $PS_i(t)$ by the Resource Governance Indicator $RGI_i$ in equation \ref{vuln} does not change the region-specific results, as seen in the supporting information, table S.\ref{si_loc}.
This suggests that $PS_i(t)$ and $RGI_i$ basically convey the same information in terms of network-based vulnerability to systemic trade-risk.
Table S.\ref{si_loc} also shows region-specific results for the case where each country is assigned the same score for $PS_i(t)=0$.
This eliminates all information on political stability of the individual countries. 
In this case the TradeRisk indicator $TR_i^{r}$ still shows higher correlations with price volatility than with the import reliance $IR_i$.

To study the robustness of the region-specific results we compare the correlation coefficients of the price volatilities of table \ref{xcorr_loc} with results from three randomized datasets, as described in section \ref{sec: rds}.
Results for randomizations are shown in the supporting information in table S.\ref{si_loc}.
For the import reliance the results do not change under any of these randomizations by construction.
Results for the In-Strength TradeRisk are  preserved under the randomizations $M^{r}_{\textrm{fix in-deg}}(t)$ and $M^{r}_{\textrm{fix in-/out-deg}}(t)$, that keep the in-degrees and both, the in- and out-degrees fixed, respectively.
This is not the case for the randomization $M^{r}_{\textrm{fix degree}}(t)$ that only preserves the average degrees of the networks.
Here we still find significant correlations between In-Strength TradeRisk and the price volatilities that are substantially lower than those for the real data.
These correlations can be attributed to the influence of the importing countries' $PS_i(t)$ values, which do not change under any of the the randomizations.
The correlations between price volatilities and both, TradeRisk and PageRank, are only significant for the the randomization scheme $M^{r}_{\textrm{fix in-/out-deg}}(t)$.
This confirms that there are substantial contributions to systemic trade-risk that can only be explained by taking the entire network of trade flows into account.

\subsection{High-risk resources}
The resource with the highest TradeRisk for the EU is beryllium.
The primary application of beryllium is in manufacturing connectors and switches for light-weight precision instruments in the aerospace and defense industries \cite{USGS}.
85\% of the world-supply of beryllium is mined in the US, most of the remainder comes from China.
Consequently the TradeRisk for the US is much lower than for the EU.
Indium has the second (third) highest TradeRisk in the EU (US).
It is essential for manufacturing liquid crystal displays.
Indium is produced almost exclusively as a byproduct of zinc mining \cite{NRC2}.
If demand for indium goes up, its availability will not necessarily increase, since this availability is largely determined by zinc economics.
The highest TradeRisk for the US is found for thallium, which is crucial for medical imaging.
Global supply of thallium is relatively constrained for the US, especially since China eliminated several tax benefits on exports of thallium in 2006 \cite{USGS}. 
We also find a high TradeRisk in the US for gallium and vanadium.
Gallium is almost exclusively produced as a byproduct of aluminium, vanadium is often a byproduct of uranium mining \cite{NRC2}.
We find a comparably high TradeRisk for tellurium in the EU (data for the US is withheld to avoid disclosing proprietary company data).
Tellurium is mined as a byproduct of copper and is critical for manufacturing solar panels \cite{USGS}.

In general we find higher TradeRisk values in the EU than in the US, see table S.\ref{details}.
The highest value of TradeRisk in the EU is 0.44 for beryllium, whereas its maximum is 0.19 for thallium in the US.

\begin{figure*}[tbp]
\begin{center}
 \includegraphics[width=130mm]{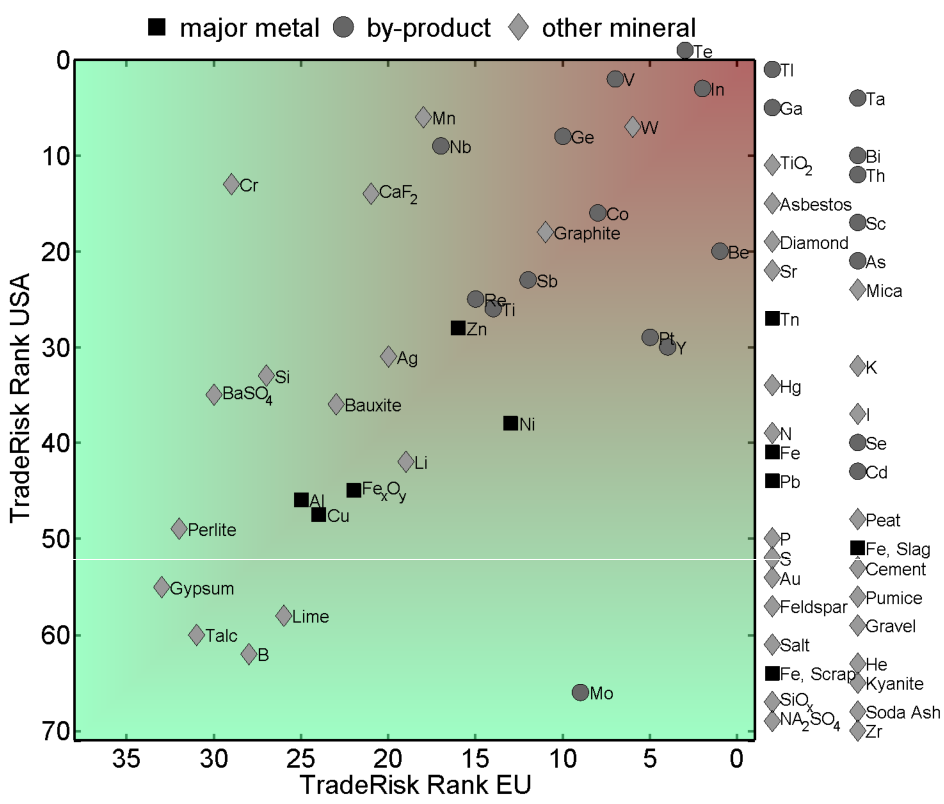}
\end{center}
 \caption{Ranks of TradeRisk in the EU and US. Each point represents a single resource. Rank 1 is given to the resource with the highest TradeRisk in the given region, rank 2 for the second highest TradeRisk, and so on. Resources where information is only available for either the EU or the US are shown outside the plot area. Major metals are shown by black boxes, minerals that are byproducts are shown as gray circles, other minerals as light-gray diamonds. It is clearly visible that those minerals that have a high TradeRisk in both regions are mined as byproducts, whereas the major metals exhibit intermediate values of TradeRisk.}
 \label{results3}
\end{figure*}

The TradeRisk Rank for individual resources is presented for the EU and US in figure \ref{results3}.
Each resource is ranked according to its TradeRisk values in the EU and US.
The lowest rank corresponds to the highest TradeRisk, the highest rank to the lowest.
Colors in figure \ref{results3} indicate whether the resources are categorized as (i) major metals (ii) byproducts of major metals, or (iii) other resources \cite{Graedel14}.
There is a tendency that those minerals that have relatively high TradeRisk in both countries tend to be mined as byproducts, whereas major metals have TradeRisks of intermediate values.

\section{Discussion}

The essence of this study is that we have shown that the international trading network of critical resources contains information that is able to explain a large fraction of the price volatility of these resources.
This information is quantified by a systemic risk measure, TradeRisk.
The correlation between TradeRisk and price volatility is substantially higher in the EU than in the US.
This finding coincides with two other facts.
First, the network-based contributions to systemic trade-risk are lower in the US, with $38\%$ of explained price volatility, than in the EU, where they explain $68\%$ of price volatility after controlling for non-network effects.
Secondly, the US systematically employs {\it lower} trade barriers for resources of high systemic trade-risk and high network-based vulnerability, whereas there are no such measures employed by the EU.
There is thus reason to assume that lower trade barriers for systemically risky resources reduce the impact of cascading supply shocks on the prices of resources. 
These findings indicate that at present there exist significant shortcomings in the risk management of non-fuel mineral resources, in particular in the EU.
This arises because systemic failure due to cumulative effects of cascading shocks on an intricately interconnected system is not taken into account. 
This is particularly salient in the light of the discovery that many of the resources most susceptible to systemic risk are only produced as byproducts and play a crucial role in industries vital to national interests.

A number of policy implications emerge from this analysis of systemic trade-risk. 
While trade in commodities or critical resources will always involve some degree of imperfect information, better monitoring and data transparency are needed to adopt a more robust approach to understanding risks that \emph{can} be foreseen by taking network effects into account. 
Policy instruments capable of mitigating systemic risk would allow decision-makers to implement measures such as strategic physical reserves and
trade regulations that mitigate market volatility while ensuring physical supply.
In financial networks it has been shown that systemic financial risk can be almost completely eliminated  by means of a so-called `systemic risk tax' \cite{RiskTax}.
This is a macroprudential regulation approach where a levy is placed on systemically risky financial transactions to offset the systemic risk increase associated with that transaction.
Motivated by this approach, it is conceivable that similar policy measures, such as a commodity risk tax, can be developed with the aim to create more resilient markets of commodities that are essential to our society.

\section{Data and methods}

\subsection{Minerals trade data}
\label{sec: mtd}
The minerals included in this study are taken from the annually published Mineral commodity summaries from the USGS \cite{USGS}.
These summaries contain information on industry structure, salient statistics, and world production and resources for 71 mineral commodities.
The summaries also list tariff codes for each mineral in the Harmonized System (HS) classification.
We collected trade data for tariff codes for each commodity as provided by the UN Comtrade data, spanning the years 2000-2012 \cite{Comtrade}.
Tariff codes that are not specific to a particular resource, such as the code 2530.90 (`other mineral substances'), are excluded.
We included all countries for which trade data for any of the minerals exists in any of these years.
This amounts to 107 countries.
Trade flows between country $i$ and country $j$ in the resource $r$ in year $t$ are recorded in the matrix $M_{ij}^{r}(t)$.
$M_{ij}^{r}(t)$ is the value of resource $r$ measured in USD that flows from country $i$ to $j$.

For each trade flow there should exist two records in the data, one for the importing country $j$, and one for the exporting country $i$.
Due to incompleteness of the data in some cases these two entries do not match. 
$M_{ij}^{r}(t)$ is defined as the larger value of these two entries.
$M_{ij}^{r}(t)$ is a time-dependent multiplex network in which the nodes correspond to countries and where each network layer is the international trade network of a given mineral $r$.
In a similar manner we construct the multiplex $K_{ij}^{r}(t)$, where each entry corresponds to the trade flow of resource $r$ from country $i$ to $j$ in year $t$, as measured in kg.
We only include trade flows that make up more than one percent of a country's imports, i.e. trade flows where $\tfrac{M_{ij}^{r}(t)}{\sum_i M_{ij}^{r}(t)}>0.01$ holds.

{\bf Price and volatility.}
The price for resource $r$ as measured in USD/kg in country $i$, $p_i(r,t)$, is obtained from the trade data as $p_i(r,t) = \tfrac{\sum_j M_{ij}^{r}(t)}{\sum_j K_{ij}^{r}(t)}$.
$p_i(r,t)$ corresponds to the average free-on-board value of resource $r$ in country $i$, that is, the transaction value of the goods and the value of services performed to deliver the goods to the border of the exporting country.
The logarithmic annual return on resource $r$ in country $i$ is $v_i(r,t) = \log \left( \tfrac{p_i(r,t)}{p_i(r,t-1)} \right)$.
The volatility of resource $r$ in country $i$, $\sigma_i^{r}$, is the standard deviation of $v_i(r,t)$, computed over the time-span $t \in \left[2000, 2012\right]$.

{\bf Total trade volume and reserves.}
The total trade volume of a resource $r$ in year $t$, $TTV(r,t)$, is the sum over all trade flows measured in kg, i.e. $TTV(r,t) = \sum_{i,j} K_{ij}^{r}(t)$.
Estimates for the available reserves of a mineral $r$, $R^{r}$, are taken from the latest estimates from the USGS \cite{USGS}.
These estimates give the future supply of identified and currently undiscovered resources that are economically extractable, taking into account also recycled resources.

\subsection{Import reliance, supply risk and trade barriers}
\label{sec: sr}

{\bf Import reliance.}
The vulnerability of a country $i$ to supply shocks in resource $r$ is strongly related to the net import reliance of $i$ on $r$.
If $i$ is a net exporter of $r$ then $i$ will be affected less by supply shocks than a country where the economy relies on imports of resource $r$.
The level of import reliance is quantified by the import reliance indicator $IR_i^{r}(t)$, for country $i$ on resource $r$ in year $t$.
For the US data on $IR_i^{r}(t)$ is provided by the USGS on an annual basis, and is defined as the imports minus exports. plus adjustments for government and industry stock exchanges \cite{USGS}.
$IR_i^{r}$ is measured as a percentage of this apparent consumption, averaged over a time-span $t \in \left[2000, 2012\right]$.
Values for the import reliance of the EU are available from the European Commission (EC) \cite{EC} for the year 2006. 

{\bf Composite supply risk.}
There exist various ways to quantify the supply risk for resources.
The (US) National Research Council (NRC) provides estimates of supply risks for 11 minerals based on import reliance, concentration of production of the resource, and substitutability \cite{NRC}.
The British Geological Survey (BGS) publishes supply risk indicators for 41 minerals taking into account their scarcity, production concentration, reserve distribution, recycling rate, substitutability, and governance aspects of the top-producing and reserve-hosting nations \cite{BGS}.
The EC releases supply risk indicators for 41 materials based on production concentration, political stability of the producing countries, as well as substitutability and recycling of the materials \cite{EC}.
We rescale each of these three lists of values such that the mineral with the highest supply risk is assigned a value of one, and the lowest supply risk is assigned a value of zero.
The composite supply risk for mineral $r$, $CSR^{r}$, is defined as the average over the rescaled supply risks provided from NRC, BGS, and the EC. 
Note that while the individual supply risk indicators are often region-specific, we regard the composite supply risk $CSR^{r}$ as a global indicator.

{\bf Trade barriers.}
The trade barriers $TB_i^{r}$ are defined as the average value of all {\it ad-valorem} equivalent tariffs per unit applied to all trade flows into region $i$ that involve resource $r$, as obtained from the MAcMap database \cite{MAcMap}.

\subsection{Randomized data sets}
\label{sec: rds}

To investigate the robustness of our results we consider three different randomizations of the trade flow matrix $M^{r}(t)$.
The randomization $M^{r}_{\textrm{fix degree}}(t)$ is obtained as a random permutation of all elements of $M^{r}(t)$.
That is, each trade flow in $M^{r}(t)$ is assigned a new importing and exporting country that is chosen at random from all countries.
$M^{r}_{\textrm{fix degree}}(t)$ has the same average degree and total trade volume as $M^{r}(t)$ but is otherwise completely randomized.
The second randomization, $M^{r}_{\textrm{fix in-deg}}(t)$, is obtained from $M^{r}(t)$ by replacing the exporting country for each trade flow by a randomly chosen country.
This randomization procedure preserves not only the average degree and the trade volume, but also the in-strength and in-degree of each country.
Network properties that involve nearest neighbors of a node, such as eigenvalues, may change under this randomization.
In the third randomization we construct the trade flow matrix $M^{r}_{\textrm{fix in-/out-deg}}(t)$ in the following way.
Let $L^{r}(t)$ be the set of links (i.e. non-zero trade flows) in $M^{r}(t)$ and $W^{r}(t)$ the corresponding set of link weights.
$M^{r}_{\textrm{fix in-/out-deg}}(t)$ is obtained by keeping $L^{r}(t)$ fixed and by replacing $W^{r}(t)$ by a random permutation of its elements.
$M^{r}(t)$ and $M^{r}_{\textrm{fix in-/out-deg}}(t)$ only differ by the volumes of the non-zero trade flows.

All results involving randomized data are averaged over 100 independent realizations of the randomization procedure.

\bibliography{literature}

\begin{thebibliography}{27}
\expandafter\ifx\csname natexlab\endcsname\relax\def\natexlab#1{#1}\fi
\expandafter\ifx\csname url\endcsname\relax
  \def\url#1{\texttt{#1}}\fi
\expandafter\ifx\csname urlprefix\endcsname\relax\def\urlprefix{URL }\fi

\bibitem[{Graedel \emph{et~al.}(2013)Graedel, Harper, Nassar \&
  Reck}]{GraedelPNAS}
Graedel, T., Harper, E., Nassar, N. \& Reck, B.~K.
\newblock On the materials basis of modern society.
\newblock \emph{Proceedings of the National Academy of Sciences}  (2013).

\bibitem[{Tang \& Xiong(2010)}]{tang2010index}
Tang, K. \& Xiong, W.
\newblock Index investment and financialization of commodities.
\newblock Tech. rep., National Bureau of Economic Research (2010).

\bibitem[{Baffes \& Haniotis(June 2010)}]{baffes2010placing}
Baffes, J. \& Haniotis, T.
\newblock Placing the 2006/08 commodity price boom into perspective.
\newblock Tech. rep., World Bank (June 2010).

\bibitem[{Billio \emph{et~al.}(2012)Billio, Getmansky, Lo \&
  Pelizzon}]{billio2012econometric}
Billio, M., Getmansky, M., Lo, A.~W. \& Pelizzon, L.
\newblock Econometric measures of connectedness and systemic risk in the
  finance and insurance sectors.
\newblock \emph{Journal of Financial Economics} \textbf{104}, 535--559 (2012).

\bibitem[{Huang \emph{et~al.}(2009)Huang, Zhou \& Zhu}]{huang2009framework}
Huang, X., Zhou, H. \& Zhu, H.
\newblock A framework for assessing the systemic risk of major financial
  institutions.
\newblock \emph{Journal of Banking \& Finance} \textbf{33}, 2036--2049 (2009).

\bibitem[{Haldane \& May(2011)}]{haldane2011systemic}
Haldane, A.~G. \& May, R.~M.
\newblock Systemic risk in banking ecosystems.
\newblock \emph{Nature} \textbf{469}, 351--355 (2011).

\bibitem[{Calvo \emph{et~al.}(2008)Calvo, Izquierdo \& Mejia}]{Calvo2008}
Calvo, G., Izquierdo, A. \& Mejia, L.-F.
\newblock Systemic sudden stops: the relevance of balance-sheet effects and
  financial integration.
\newblock \emph{Inter-American Development Bank. Research Department Working
  Papers}  (2008).

\bibitem[{Grubesic \emph{et~al.}(2008)Grubesic, Matisziw, Murray \&
  Snediker}]{grubesic2008comparative}
Grubesic, T.~H., Matisziw, T.~C., Murray, A.~T. \& Snediker, D.
\newblock Comparative approaches for assessing network vulnerability.
\newblock \emph{International Regional Science Review} \textbf{31}, 88--112
  (2008).

\bibitem[{Simonsen \emph{et~al.}(2008)Simonsen, Buzna, Peters, Bornholdt \&
  Helbing}]{simonsen2008transient}
Simonsen, I., Buzna, L., Peters, K., Bornholdt, S. \& Helbing, D.
\newblock Transient dynamics increasing network vulnerability to cascading
  failures.
\newblock \emph{Physical review letters} \textbf{100}, 218701 (2008).

\bibitem[{Battiston \emph{et~al.}(2012)Battiston, Puliga, Kaushik, P \&
  G}]{Battiston12}
Battiston, S., Puliga, M., Kaushik, R., P, T. \& G, C.
\newblock Debtrank: Too central to fail? financial networks, the FED and
  systemic risk.
\newblock \emph{Scientific Reports} \textbf{2} (2012).

\bibitem[{Thurner \& Poledna(2013)}]{ThurnerSRT}
Thurner, S. \& Poledna, S.
\newblock DebtRank-transparency: Controlling systemic risk in financial
  networks.
\newblock \emph{Scientific Reports}  (2013).

\bibitem[{Lee \emph{et~al.}(2012)Lee, Preston, Kooroshy, Bailey \&
  Lahn}]{Chatham}
Lee, B., Preston, F., Kooroshy, J., Bailey, R. \& Lahn, G.
\newblock \emph{Resources futures}, vol.~14 (Citeseer, 2012).

\bibitem[{Erdmann \& Graedel(2011)}]{erdmann2011criticality}
Erdmann, L. \& Graedel, T.~E.
\newblock Criticality of non-fuel minerals: a review of major approaches and
  analyses.
\newblock \emph{Environmental science \& technology} \textbf{45}, 7620--7630
  (2011).

\bibitem[{Gunn \emph{et~al.}(2014)Gunn, Graedel \& Espinoza}]{Graedel14}
Gunn, G., Graedel, T. \& Espinoza, L.~T.
\newblock \emph{Metal resources, use and criticality} (Wiley, 2014).

\bibitem[{(US)(2014)}]{USGS}
(US), G.~S.
\newblock Mineral Commodity Summaries, 2014.
\newblock Tech. rep., United States Geological Survey (2014).

\bibitem[{Graedel \emph{et~al.}(2012)}]{Graedel}
Graedel, T. \emph{et~al.}
\newblock Methodology of metal criticality determination.
\newblock \emph{Environmental science \& technology} \textbf{46}, 1063--1070
  (2012).

\bibitem[{Kaufmann \emph{et~al.}(September 2010)Kaufmann, Kraay \&
  Mastruzzi}]{WGI}
Kaufmann, D., Kraay, A. \& Mastruzzi, M.
\newblock The Worldwide Governance Indicators: Methodology and Analytical
  Issues.
\newblock Tech. rep., World Bank (September 2010).

\bibitem[{Institute(2013)}]{RGI}
Institute, R.~W.
\newblock The 2013 Resource Governance Index.
\newblock Tech. rep., Revenue Watch (2013).

\bibitem[{Newman(2010)}]{Newman}
Newman, M.
\newblock \emph{Networks: an introduction} (Oxford University Press, 2010).

\bibitem[{Wang \emph{et~al.}(2003)Wang, Chakrabarti, Wang \&
  Faloutsos}]{Wang03}
Wang, Y., Chakrabarti, D., Wang, C. \& Faloutsos, C.
\newblock Epidemic spreading in real networks: An eigenvalue viewpoint.
\newblock In \emph{Reliable Distributed Systems, 2003. Proceedings. 22nd
  International Symposium on}, 25--34 (IEEE, 2003).

\bibitem[{NRC(2012)}]{NRC2}
The Role of the Chemical Sciences in Finding Alternatives to Critical
  Resources: A Workshop Summary.
\newblock Tech. rep., National Research Council (US), Washington, DC (2012).

\bibitem[{Poledna \& Thurner(2014)}]{RiskTax}
Poledna, S. \& Thurner, S.
\newblock Elimination of systemic risk in financial networks by means of a
  systemic risk tax.
\newblock \emph{arXiv:1401.8026}  (2014).

\bibitem[{Com(2010)}]{Comtrade}
United Nations commodity trade statistics database.
\newblock Tech. rep., Comtrade, UN (2010).

\bibitem[{Commission(2014)}]{EC}
Commission, E.
\newblock DG Enterprise and Industry, Critical raw materials for the EU.
\newblock Tech. rep., European Commission (2014).

\bibitem[{NRC(2008)}]{NRC}
Minerals, critical minerals, and the US economy.
\newblock Tech. rep., National Research Council (US). Committee on Critical
  Mineral Impacts on the US Economy (2008).

\bibitem[{Survey(2012)}]{BGS}
Survey, B.~G.
\newblock Risk List 2012.
\newblock Tech. rep., British Geological Survey (2012).

\bibitem[{Bou{\"e}t \emph{et~al.}(2004)Bou{\"e}t, Decreux, ́, Jean \&
  Debucquet}]{MAcMap}
Bou{\"e}t, A., Decreux, Y., ́, L.~F., Jean, S. \& Debucquet, D.~L.
\newblock \emph{A consistent, ad-valorem equivalent measure of applied
  protection across the world: The MAcMap-HS6 database} (Centre d'{\'e}tudes
  prospectives et d'informations internationales (CEPII), 2004).

\end{thebibliography}

\begin{table*}[tbp]
\caption{Network-based properties, supply risk, and indicators obtained from trade data for 71 non-fuel mineral resources. For each resource $r$ the values of the composite supply risk $CSR^{r}$, the exponential scarcity $S^{r}$, total trade volume $TTV^{r}$, average degree $\bar k^{r}$, the largest eigenvalue $\lambda^{r}$ (inverse resilience), the size of the largest strongly connected component $SCC^{r}$, and the TradeRisk for EU and US, $TR_{EU}^{r}$ and $TR_{US}^{r}$, are shown. }
\label{details}
\footnotesize
\begin{tabular}{l | l l l l l l | l l}
\hline
               &               &               &               & global              &               &              &     local          &               \\
Mineral $r$    &  $CSR^{r}$     &    exp$(S^{r})$&  $TTV^{r}$    &  $\bar k^{r}$& $\lambda^{r}$&$SCC^{r}$     & $TR_{EU}^{r}$  &$TR_{US}^{r}$  \\
\hline
Aluminium      &  0.06         &               & 5.3  $10^9$   & 6.7           &   0.46        &  0.20         & 0.021         &   0.0037      \\
Antinomy       & 0.71          & 3.7 $10^{-5}$ &6.7 $10^7$     & 4.8           &  0.57         &  0.16         & 0.076         &   0.029       \\               
Arsenic        & 0.63          & 3.4 $10^{-6}$ &3.0 $10^6$     & 2.3           &  0.33         &  0.082        &               &   0.041       \\               
Asbestos       &               & 4.8 $10^{-5}$ &5.3   $10^8$   & 1.6           &  0.54         &  0.089        &               &   0.060       \\               
Barite         & 0.53          & 0.0016        & 5.5  $10^8$   & 5.0           &  0.41         &  0.13         & 0.0047        &   0.010       \\               
Bauxite        & 1             & 5.9 $10^{-4}$ & 1.6  $10^{10}$& 5.2           &  0.42         &  0.099        & 0.022         &   0.0097      \\               
Beryllium      & 0.49          & 2.7 $10^{-5}$ & 4.0  $10^5$   & 1.6           &  0.25         &  0.063        & 0.44          &   0.045       \\               
Bismuth        & 0.90          & 2.6 $10^{-5}$ & 8.2  $10^6$   & 3.6           &  0.45         &  0.11         &               &   0.082       \\               
Boron          & 0.10          & 0.0041        & 8.6  $10^8$   & 6.0           &  0.46         &  0.093        & 0.021         &    0          \\               
Cadmium        & 0.46          & 2.4 $10^{-5}$ & 1.2  $10^7$   & 3.4           &  0.42         &  0.16         &               &   0.0052      \\               
Cement         &               &               & 8.1  $10^{10}$& 6.1           &   0.45        &  0.30         &               &  4.5 $ 10^{-4}$\\              
Chromium       & 0.24          & 0.0056        & 2.7  $10^9$   & 5.3           &   0.54        &  0.11         &  0.023        &  0.071        \\               
Cobalt         & 0.42          & 1.9 $10^{-5}$ & 1.4  $10^8$   & 5.9           &   0.31        &  0.083        &  0.11         &  0.060        \\               
Copper         & 0.0062        & 0.0089        & 6.1  $10^9$   & 5.7           &   0.52        &  0.18         &  0.027        &  0.0034       \\               
Diamond        & 0.37          & 0.0014        & 1.1  $10^6$   & 5.1           &   0.50        &  0.10         &               &  0.053        \\               
Feldspar       & 0.021         &               & 2.9  $10^9$   & 3.4           &   0.63        &  0.16         &               &4.8 $ 10^{-5}$ \\               
Fluorspar      & 0.40          & 8.8 $10{-5}$  & 2.1  $10^7$   & 3.7           &   0.39        &  0.12         &  0.024        & 0.070         \\               
Gallium        & 0.60          &               & 8.8  $10^7$   & 4.9           &   0.43        &  0.11         &               & 0.094         \\               
Germanium      & 0.65          &               & 2.5  $10^7$   & 4.9           &   0.36        &  0.12         &   0.097       & 0.086         \\               
Gold           & 0.27          & 3.3 $10^{-4}$ & 1.8  $10^7$   & 4.2           &   0.42        &  0.22         &               &2.5 $10^{-4}$  \\               
Graphite       & 0.49          &               & 3.4  $10^8$   & 4.6           &   0.41        &  0.089        &   0.082       & 0.055         \\               
Gypsum         & 0.049         &               & 1.1  $10^{10}$& 3.6           &   0.47        &  0.17         &  6.0 $10^{-4}$&2.4 $10^{-4}$  \\               
Helium         &               &   0.0031      & 2.3  $10^7$   & 4.7           &   0.51        &  0.080        &               &  0            \\               
Titanium       & 0.14          & 3.9 $10^{-4}$ &2.6   $10^9$   & 3.3           &  0.44         &  0.12         &  0.12         &   0.022       \\               
Indium         & 0.57          &               &5.9   $10^6$   & 3.7           &  0.42         &  0.12         &  0.27         &   0.10        \\               
Iodine         &               & 1.5 $10^{-6}$ &1.1   $10^7$   & 5.6           &  0.35         &  0.095        &               &   0.0095      \\               
Iron Ore       & 0.11          & 2.3           & 3.9  $10^{11}$& 2.9           &  0.58         &  0.12         &  0.059        &    0.0046     \\               
Iron and Steel &               &               & 1.0  $10^{10}$& 5.5           &  0.50         &  0.13         &               &    0.0064     \\               
Iron Scrap     &               &               & 4.7  $10^{10}$& 6.0           &  0.50         &  0.27         &               &   0           \\               
Iron Slag      &               &               & 7.2  $10^{9 }$& 3.2           &  0.53         &  0.23         &               &  9.8 $10^{-4}$\\               
Kyanite        &               &               & 2.6  $10^8$   & 3.8           &  0.37         &  0.10         &               &  0            \\               
Lead           & 0.37          &     0.014     & 1.3  $10^9$   & 4.5           &  0.58         &  0.19         &               &  0.0047       \\               
Lime           & 0.13          &               & 1.9  $10^9$   & 4.3           &  0.41         &  0.20         &  0.014        & 1.0 $10^{-5}$ \\               
Lithium        & 0.31          & 2.3 $10^{-6}$ & 2.9  $10^7$   & 4.0           &  0.41         &  0.037        &  0.031        & 0.0062        \\               
Manganese      & 0.33          &     0.010     & 5.9  $10^9$   & 5.9           &  0.38         &  0.13         &  0.046        & 0.091         \\               
Mercury        &               & 2.2 $10^{-5}$ & 2.1  $10^6$   & 3.2           &  0.35         &  0.12         &               & 0.011         \\               
Mica (Sheet)   & 0.83          &               & 3.8  $10^7$   & 6.5           &  0.39         &  0.085        &               & 0.028         \\               
Molybdenum     & 0.45          &     0.0061    & 6.7  $10^7$   & 5.5           &  0.48         &  0.12         &   0.12        & 0             \\               
Nickel         & 0.20          & 1.0 $10^{-5}$ & 7.5  $10^8$   & 5.3           &  0.33         &  0.072        &   0.063       & 0.0089        \\               
Niobium        & 0.63          & 2.5 $10^{-5}$ & 1.1  $10^8$   & 3.9           &  0.42         &  0.095        &   0.049       & 0.084         \\               
Nitrogen       &               &               & 1.6  $10^{10}$& 6.7           &  0.54         &  0.17         &               & 0.0075        \\               
Peat           &               & 3.8 $10^{-4}$ &  4.5 $10^9$   & 5.6           &  0.31         &  0.11         &               & 0.0033        \\               
Perlite        & 0.028         &     0.012     &  1.2 $10^9$   & 4.3           &  0.55         &  0.12         &   0.0075      & 0.0023        \\               
Phosphate Rock &               & 1.9 $10^{-4}$ &  1.3 $10^{10}$& 2.5           &  0.59         &  0.12         &               & 0.0010        \\               
Platinum       & 0.80          & 9.6 $10^{-8}$ &  6.4 $10^6$   & 4.2           &  0.37         &  0.066        &   0.11        & 0.018         \\               
Potash         &               &     0.0037    &  2.2 $10^{10}$& 5.5           &  0.42         &  0.12         &               & 0.013         \\               
Pumice         &               &               &  9.8 $10^8$   & 6.1           &  0.49         &  0.10         &               & 5.8 $10^{-5}$ \\               
Quartz         &               &               &  4.6 $10^9$   & 8.4           &  0.37         &  0.15         &               &  0            \\               
Rare Earth     &  1            & 3.9 $10^{-7}$ &  5.4 $10^7$   & 4.6           &  0.43         &  0.050        &   0.093       &  0.016        \\               
Rutile         &               & 2.4 $10^{-6}$ &  5.9 $10^6$   & 3.7           &  0.42         &  0.12         &               &  0.082        \\               
Rhenium        & 0.26          &               &  2.6 $10^9$   & 3.3           &  0.44         &  0.12         & 0.12          &  0.023        \\               
Salt           &               &               & 2.5  $10^{10}$& 6.7           &  0.43         &  0.21         &               &5.8 $10^{-11}$ \\               
Sand and Gravel&               &               & 5.8  $10^{10}$& 4.8           &  0.50         &  0.22         &               & 1.4 $10^{-6}$ \\               
Scandium       &               &               & 3.2  $10^7$   & 4.1           &  0.47         &  0.046        &               &  0.056        \\ 
Selenium       &  0.54         & 6.5 $10^{-5}$ & 7.8  $10^6$   & 3.9           &  0.37         &   0.15        &               &  0.0069       \\               
Tellurium      &  0.10         &               & 1.5  $10^6$   & 3.2           &  0.37         &   0.14        &  0.21         &               \\                             
Silicon        &  0.041        &               & 6.0  $10^8$   & 6.3           &  0.39         &   0.11        &  0.013        &  0.013        \\               
Silver         &  0.20         & 8.0 $10^{-6}$ & 4.2  $10^6$   & 4.8           &  0.37         &   0.077       &  0.035        &  0.015        \\               
Soda Ash       &               & 2.2 $10^{-4}$ & 5.3  $10^9$   & 4.2           &  0.51         &   0.13        &               &  0            \\               
Sodium Sulfate &               &               & 1.8  $10^7$   & 2.7           &  0.41         &   0.055       &               &  0            \\               
Strontium      &  0.83         & 1.4 $10^{-5}$ & 9.4  $10^7$   & 3.1           &  0.41         &   0.092       &               &  0.035        \\               
Sulfur         &               &               & 1.1  $10^{10}$& 5.4           &  0.45         &   0.20        &               & 9.8 $10^{-4}$ \\               
Talc           &  0.037        &               & 1.6  $10^9$   & 6.0           &  0.37         &   0.063       &   0.0055      & 8.0 $10^{-7}$ \\               
Tantalum       &  0.37         & 8.3 $10^{-4}$ & 8.3  $10^7$   & 3.9           &  0.39         &   0.15        &               &   0.10        \\               
Thallium       &               &               & 9.9  $10^5$   & 1.8           &  0.41         &   0.15        &               &   0.19        \\               
Thorium        &  0.63         & 2.2 $10^{-6}$ & 3.1  $10^6$   & 0.91          &  0.13         &   0.041       &               &   0.073       \\               
Tin            &  0.46         & 4.1 $10^{-5}$ & 1.9  $10^8$   & 4.8           &  0.55         &   0.11        &               &   0.021       \\               
Tungsten       &  0.68         & 3.5 $10^{-6}$ & 1.2  $10^7$   & 4.3           &  0.47         &   0.12        &  0.099        &   0.0089      \\               
Vanadium       &  0.31         &     0.0023    & 3.2  $10^7$   & 4.4           &  0.42         &   0.12        &  0.11         &   0.12        \\               
Zinc           &  0.076        &     0.020     & 5.0  $10^9$   & 6.1           &  0.58         &   0.17        &  0.062        &   0.020       \\               
Zirconium      &  0.27         &     0.0037    & 2.5  $10^8$   & 5.3           &  0.40         &   0.15        &               &        0      
\end{tabular}
\end{table*}

\begin{table*}[bp]
\caption{Global properties of the randomized trade networks $M^{r}_{\textrm{fix degree}}(t)$. There are no significant correlations between the largest eigenvalue $\lambda^{r}$ and any of the parameters for the randomized data. This confirms that the observed relations between scarcity and $\lambda^{r}$,and between  supply risk and $\lambda^{r}$,  are indeed the result of non-trivial network effects. The correlations involving the size of the largest component $SCC^{r}$ are also weaker than in the randomized data.}
\label{si_glob}
\begin{tabular}{l l l l}
\hline
 & & $M^{r}_{\textrm{fix degree}}(t)$ & \\
Correlation between &    $CSR^{r}$ &  $S^{r}$  & $TTV^{r}$ \\
\hline
Largest Eigenvalue $\lambda^{r}$ 		& $-0.12$		& $0.18 $		& $0.05$ \\
Size SCC $SCC^{r}$ 						& $-0.21$		& $0.30^{*}$	& $0.03$ \\
\hline
\end{tabular}
\end{table*}

\begin{table*}[tbp]
\caption{Pearson correlation coefficients between various price volatilities with TradeRisk, import reliance, PageRank, and in-strength for the EU and US and for several variants of the calculations. The columns `no PS' show results for the case where the influence of political stability of the importing countries is neglected ($PS_i=0$ for all $i$). For `RGI' the values for the political stability $PS_i(t)$ have been replaced by values for the Resource Governance Indicator $RGI_i$. Results for the three different randomizations $M^{r}_{\textrm{fix degree}}(t)$,  $M^{r}_{\textrm{fix in-deg}}(t)$, and  $M^{r}_{\textrm{fix in-/out-deg}}(t)$ are also shown. Results for $IR_i^r$ are independent of the network and are shown only for comparison. The significant correlations between In-Strength TradeRisk and the price volatilities are driven by the import reliance and to some extent also by the political stability of the neighboring countries. In some cases the correlation between $TR_i^{str,r}$ and the price volatilities is even slightly higher than for the data, which suggests that countries actively avoid to trade critical resources politically unstable countries.}
\label{si_loc}
\begin{tabular}{l l l | l l | l l | l l | l l }
\hline
 & `no PS' & & `RGI' & &  $M^{r}_{\textrm{fix degree}}(t)$ &  &  $M^{r}_{\textrm{fix in-deg}}(t)$ & &  $M^{r}_{\textrm{fix in-/out-deg}}(t)$ & \\
Correlation between &    $\sigma_{EU}^{r}$  &  $\sigma_{US}^{r}$  &    $\sigma_{EU}^{r}$  &  $\sigma_{US}^{r}$&    $\sigma_{EU}^{r}$  &  $\sigma_{US}^{r}$&    $\sigma_{EU}^{r}$  &  $\sigma_{US}^{r}$&    $\sigma_{EU}^{r}$  &  $\sigma_{US}^{r}$    \\
\hline
TradeRisk $TR_i^{r}$						& $0.58^{***}$	& $0.55^{***}$	& $0.71^{***}$  & $0.56^{***}$	& $-0.02$		& $0.07$		& $0.26$		& $0.22$		& $0.57^{**}$	& $0.54^{***}$	\\
Import Reliance $IR_i^{r}$				& $0.48^{**}$   & $0.51^{***}$	& $0.48^{**}$   & $0.51^{***}$	& $0.48^{**}$   & $0.51^{***}$	& $0.48^{**}$   & $0.51^{***}$	& $0.48^{**}$   & $0.51^{***}$	\\
PageRank $PR_i^{r}$					& $0.57^{***}$	& $0.30^*$		& $0.57^{***}$	& $0.46^{***}$	& $-0.21$		& $-0.06$		& $-0.10$		& $-0.04$		& $0.41^{**}$	& $0.37^*$		\\
In-Strength TradeRisk $TR_i^{str,r}$ 	& $0.48^{**}$	& $0.51^{***}$	& $0.48^{**}$	& $0.52^{***}$	& $0.43^*$		& $0.51^{***}$	& $0.48^{**}$	& $0.51^{***}$	& $0.45^{*}$	& $0.51^{***}$	\\
\hline
\end{tabular}
\end{table*}
\end{document}